\documentclass[aps,twocolumn,nofootinbib]{revtex4}
\usepackage{graphicx}
\usepackage{epsfig}
\usepackage{color}

\usepackage{calc}
\usepackage{ifthen}

{
{\newcommand{\gsim}{\mbox{\raisebox{-.6ex}{~$\stackrel{>}{\sim}$~}}}
{
\newcommand{\be}{\begin{equation}}
\newcommand{\ee}{\end{equation}}
\newcommand{\bea}{\begin{eqnarray}}
\newcommand{\eea}{\end{eqnarray}}

\def\GeV{{\rm \ GeV}}

\def\TeV{{\rm \ TeV}}

 \def\gae{\; ^{>}_{\sim} \;}

\begin{document}
\rightline{KEK-Cosmo-121,KEK-TH-1642}

\title{Constraining the co-genesis of Visible and Dark Matter with AMS-02 and Xenon-100}
\author{Kazunori Kohri~$^{1}$}
\author{Narendra Sahu~$^{2}$}
\affiliation{$^{1}$~Cosmophysics Group, Theory Centre, IPNS, KEK, Tsukuba 305-0801, Japan\\
The Graduate University for Advanced Study (Sokendai), Tsukuba 305-0801, Japan\\
$^2$~Department of Physics, Indian Institute of Technology Hyderabad, Yeddumailaram 502205, AP, India}

\begin{abstract}
We study  a non-thermal scenario in a two-Higgs doublet extension of the standard model (SM), augmented by an $U(1)_{\rm B-L}$ 
gauge symmetry. In this set up, it is shown that the decay product of a weakly coupled scalar field just above the electroweak 
scale can generate visible and dark matter (DM) simultaneously. The DM is unstable because of the broken $B-L$ symmetry. 
The lifetime of DM ($\approx 5\times 10^{25}$ sec) is found to be much longer than the age of the Universe, and 
its decay to the SM leptons at present epoch can explain the positron excess observed at the AMS-02. The relic abundance and 
the direct detection constraint from Xenon-100 can rule out a large parameter space just leaving the $B-L$ breaking scale 
around $\approx 2 - 4$ TeV.   

\end{abstract}

\maketitle

\section{Introduction} \label{section-1}

The observed cosmic ray anomalies at PAMELA~\cite{PAMELA_positron, PAMELA_ep},
Fermi~\cite{Fermi_ep,Fermi_positron}, H.E.S.S.~\cite{hess_ep} and recently at AMS-02~\cite{ams,ams_ep} (see also
\cite{cosmicray_expts}) conclusively hint towards a primary source of positron in our 
Galaxy~\footnote{ In
  fact it has been shown earlier that there is a clean excess of
  absolute positron flux in the cosmic rays at an energy $E\gsim 50$
  GeV~\cite{BMS_09}, even if the propagation
  uncertainty~\cite{uncertainty} in the secondary positron flux is
  added to the Galactic background.}. This gives rise enough
motivation to consider a particle physics based dark matter (DM) models,
such as annihilation~\cite{annihilation, Hamaguchi:2009jb,Kopp:2013eka,
DeSimone:2013fia,Yuan:2013eja,Jin:2013nta} or
decay~\cite{decay,Kohri:2009yn,Hamaguchi:2009jb,Ibe:2013nka,Jin:2013nta,Kajiyama:2013dba,Ibe:2013jya,Feng:2013vva} of DM, as the
origin of positron excess in the cosmic rays~\footnote{For
  astrophysical origins, see Ref.~\cite{astro-orig,Yuan:2013eja,Yuan:2013eba,Cholis:2013psa,Linden:2013mqa} and
  references therein.}. 

At present, the relic abundance of DM: $\Omega_{\rm DM} h^2\sim 0.12 $, is well measured by the Planck
satellite~\cite{Planck}. However, the mechanism that provides its relic abundance is not yet 
established.  Moreover, the origin of tiny amount of visible matter in the Universe which is in the form of
baryons with $\Omega_{\rm b}h^2 \sim 0.022$ arising from a baryon asymmetry: $n_B/n_\gamma 
\sim 6.15 \times 10^{-10}$, has been established by the Planck~\cite{Planck} and the big-bang nucleosynthesis (BBN) 
measurements~\cite{pdg}. The fact that the DM abundance is about a factor of 5 with respect to the baryonic one might
hint towards a common origin behind their genesis.

In fact, both baryon and DM abundances could be produced at the end of inflation, whose origin is usually linked 
to a scalar field called inflaton~\cite{infl-rev}. A visible sector inflaton which carries the Standard Model (SM) charges~\cite{visible}
can naturally create a weakly interacting DM, as it happens in the case of Minimal Supersymmetric SM scenarios, see~\cite{vis-dark}. 
However if the inflaton belongs to a hidden sector, such a SM singlet inflaton, which might as well couple to other hidden sectors, then it becomes 
 a challenge to create the right abundance for both DM and the visible matter. 

In this paper we will consider a simple example of any generic hidden sector inflaton, which first decays into scalar fields 
charged under a $U(1)_{B-L}$ gauge group. The subsequent decay of these scalar fields to DM and SM charged leptons generate 
asymmetry in the visible and DM sectors, which has to be matched with the observed data~\cite{Planck}. The stabilty 
to DM is provided by the $B-L$ gauge symmetry. We assume that all the above phenomena happens in a non-thermal scenario right above 
the electroweak scale.

If we assume that $B-L$ is broken above the TeV scale, then the resulting DM lifetime comes out to be longer than the age of the universe,
i.e. $\approx 5\times 10^{25}$ sec, and it's decay into charged leptons can explain the rising positron spectrum as shown by the AMS-02 data,
provided that the DM mass is around $1$~TeV. Furthermore, we are able to put constraints on the model parameters by the direct detection experiments,
such as Xenon-100~\cite{Aprile:2012nq}. The null-detetction of DM at Xenon-100 constraints the $B-L$ breaking scale to be around $2 - 4$ TeV. The 
model can be further constrained by the LHC if there is a discovery of an extra $Z'$ gauge boson.

The paper is organized as follows. In section-\ref{section-2}, we briefly discuss the model. In section-\ref{section-3}, 
we provide the mechanism of generating visible and DM simultaneously in a non-thermal set-up. In section-\ref{section-4} 
we discuss positron anomalies from a decaying DM. In section -\ref{section-5}, we discuss compatibility of 
the DM with the direct detection limits. In section-\ref{section-6}, we conclude our main results.

\section{The Model} \label{section-2}

The positron excess seen in PAMELA~\cite{PAMELA_positron,PAMELA_ep}, Fermi~\cite{Fermi_ep,Fermi_positron}, AMS-02~\cite{ams,ams_ep} experiments hint towards a 
leptophilic origin of the DM~\cite{leptophilic,Kohri:2009yn}. A simple non-supersymmetric origin of this DM can be explained in 
a two Higgs doublet extension of the SM with an introduction of an $U(1)_{\rm B-L}$ gauge symmetry~\cite{sahu&sarkar,Kohri:2009yn}. 
We also add three singlet fermions $N_L (1,0,-1)$, $\psi_R (1,0,-1)$ and $S_R (1,0,-1)$ per generation, where the numbers 
inside the parentheses indicate their quantum numbers under the gauge group $SU(2)_L \times U(1)_{Y}\times U(1)_{\rm B-L}$. We 
need to check the axial-vector anomaly~\cite{anomaly_condition}, which requires the following conditions to be satisfied for 
its absence:
\begin{eqnarray*}
SU(3)_C^2 ~ U(1)_{\rm B-L}  &:&  3\left[2 \times \frac{1}{3}- \frac{1}{3}-\frac{1}{3}\right] = 0 \nonumber\\
SU(2)_L^2 ~ U(1)_{\rm B-L} &:&  2 \left[  \frac{1}{3}\times 3 + (-1) \right]  = 0 \nonumber\\
U(1)_Y^2 ~ U(1)_{\rm B-L}  &:&  3 \left[ 2 \times \left( \frac{1}{3} \right)^2 \times \frac{1}{3} \right] \nonumber\\
&-& 3 \left[  \left( \frac{4}{3} \right)^2 \times \frac{1}{3} + \left(\frac{-2}{3} \right)^2 \times \frac{1}{3}\right]\nonumber\\ 
&+& \left[2 (-1)^2 (-1) - 1  (-2)^2  (-1) \right]=0\nonumber\\
U(1)_Y ~ U(1)_{\rm B-L}^2 &:& 3 \left[ 2 \times \frac{1}{3} \times \left(\frac{1}{3} \right)^2 \right] \nonumber\\
&-& \left[  \frac{4}{3} \times \left(\frac{1}{3} \right)^2 + \left(\frac{-2}{3} \right)\times \left( \frac{1}{3} \right)^2 \right] \nonumber\\
&+& \left[  2 (-1)  (-1)^2 - 1 (-2) (-1)^2  \right]=0 \nonumber\\
 U(1)_{\rm B-L}^3  &:&  3 \left[ 2 \times \left(\frac{1}{3}\right)^3 - \left( \frac{1}{3} \right)^3- \left( \frac{1}{3} \right)^3\right] \nonumber\\
&+&  \left[ 2\times (-1)^3 - (-1)^3 \right]  \nonumber\\
&+&  \left[ (-1)^3- (-1)^3 -(-1)^3\right]=0 \nonumber
\end{eqnarray*}
where the number 3 in front is the color factor. Thus the model is shown to be free from $B-L$ anomaly and hence can be gauged by introducing 
an extra gauge boson $Z'$. Since $N_L$ is a singlet under $SU(2)_L$, and it does not carry any charge under $U(1)_Y$, its electromagnetic charge 
is zero. As a result the lightest one can be a viable candidate of the DM. The stability to DM is provided by the gauged $B-L$ symmetry. 

\begin{figure}
\begin{center}
\epsfig{file=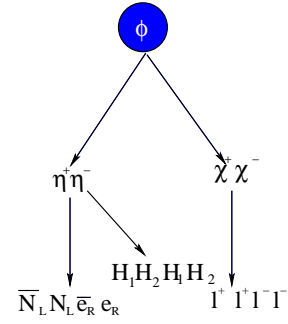, width=0.2\textwidth}
\caption{Decay of hidden sector inflaton to SM degrees of freedom through $\eta$ and $\chi$ fields.}
\label{fig-1}
\end{center}
\end{figure}

However, we also add two massive charged scalars: $\eta^- (1,-2,0)$ and $\chi^-(1,-2,-2)$ in the particle spectrum such that their 
interaction in the effective theory breaks lepton number by two units and hence introduces a prolonged lifetime for the lightest $N_L$, 
which is the candidate for DM. As we show later the extremely slow decay of DM can explain the positron excess observed 
at PAMELA~\cite{PAMELA_positron}, Fermi~\cite{Fermi_positron} and recently at AMS-02~\cite{ams}. Furthermore, we assume that these particles are 
produced non-thermally from the cascade decay of the hidden sector inflaton field $\phi (1,0,0)$ just above the EW scale as pictorially 
depicted in Fig. \ref{fig-1}. The particle content and their quantum numbers are summarised in table~\ref{table-1}.
\begin{table}[h]\label{table-1}
\begin{center}
\caption{Particle content and their quantum numbers.\label{table-1}}
\begin{tabular}{|c|c|c|c|}\hline
Particle & $SU(2)_L\times U(1)_Y$ &  $U(1)_{B-L}$ & Mass range\\ \hline
$\ell_L$  & (2,-1) & -1 & MeV to GeV \\[2mm] \hline
$\ell_R^-$ & (1,-2) & -1 & MeV to GeV\\[2mm] \hline
$H_1$, $H_2$  & (2,1) & 0 & 100 GeV $\to {\cal O}({\rm TeV})$\\[2mm] \hline
$\phi$ & (1,0) & 0 & ${\cal O}( 10^3 {\rm TeV})$\\[2mm]\hline
$\chi^-$   & (1,-2) & -2  & ${\cal O}(10^3  {\rm TeV})$\\[2mm] \hline
$\eta^-$ & (1,-2) & 0 & ${\cal O}( 10^3 {\rm TeV}) $ \\[2mm] \hline
$N_L$ & (1,0) & -1 & ${\cal O}({\rm TeV})$\\[2mm] \hline
$\psi_R$, $S_R$ & (1,0) & -1 & ${\cal O}({\rm TeV})$ \\[2mm]\hline
\end{tabular}
\end{center}
\end{table}


%

The main interactions are given by the effective Lagrangian:
\begin{eqnarray}
{\cal L}_{\rm eff} &\supseteq  & \frac{1}{2}(M_N)_{\alpha \beta}\overline{(N_{\alpha L})^c}N_{\beta L} 
+\frac{1}{2}(M_\psi)_{\alpha \beta}\overline{(\psi_{\alpha R})^c}\psi_{\beta R} \nonumber\\
&+& \frac{1}{2}(M_S)_{\alpha \beta}\overline{(S_{\alpha R})^c}S_{\beta R} + (g_S)_{\alpha \beta}\left( 
\overline{S_{\alpha R}}H \ell_{\beta L} \right) \nonumber\\
&+& (g_\psi)_{\alpha \beta} \left( \overline{\psi_{\alpha R}} H \ell_{\beta L}\right)
+ \mu \eta H_1 H_2 +m^2 \eta^\dagger \chi \nonumber\\
&+& h_{\alpha \beta} \eta^\dagger \overline{N_{\alpha L}} \ell_{\beta R} +
f_{\alpha \beta} \chi^\dagger \ell_{\alpha L} \ell_{\beta L}+ h.c.\,
\label{Lagrangian}
\end{eqnarray} 
where 
\begin{equation}
m^2=\mu' v_{\rm B-L},~~M_i=F_i v_{\rm B-L}\,,
\end{equation}
 with $``v_{\rm B-L}"$ is the vacuum expectation value (vev) of the $U(1)_{\rm B-L}$ breaking scalar field which carries 
$B-L$ charges by two units and $F_i$ is the coupling between $B-L$ breaking scalar field and the singlet fermions. In 
Eq.~(\ref{Lagrangian}), $H_1$, $H_2$ are two Higgs doublets and $\ell_L (2,-1,-1)$, $\ell_R(1,-2,-1)$ are SM lepton 
doublet and singlet respectively. 

We demand $M_i=F_i v_{\rm B-L}$, with $i=N,S,\psi$, to be of the order of TeV scale in order to explain the cosmic ray anomalies 
as discussed in section \ref{section-4}. Since the interactions of $S$ and $\psi$ break $B-L$ by two units, the neutrino mass, 
after electroweak phase transition, can be generated via the dimension five operators: $\ell \ell HH/M_S$ and $\ell_L \ell_L H H/M_\psi$ 
and is given by: 
\begin{equation}
M_\nu=\frac{g_S^2 \langle H \rangle^2 }{M_S} + \frac{g_\psi^2 \langle H \rangle^2 }{M_\psi}\,. 
\end{equation} 
Taking $M_S, M_\psi \sim {\cal O} (\rm TeV)$, the sub-eV neutrino mass imply $g_S, g_\psi \sim {\cal O} (10^{-5})$. Therefore, the 
decay of $S$ and $\psi$ can not produce any lepton asymmetry even though their interactions break $B-L$ by two units. Moreover, the 
number density of these particles are Boltzmann suppressed as the reheat temperature is around 100 GeV.   

As we will show in section (\ref{section-3}), the lepton number conserving decay: $\eta \to N_L + \ell_R$ generates visible 
and DM ($N_L$) simultaneously. However, note that the interaction between $\eta$ and $\chi$ violates the lepton number 
by two units. Therefore, the DM is no more stable and decays slowly to SM fields. Since the DM carry a net leptonic charge, 
it only decays to leptons without producing any quarks. As we will discuss in section (\ref{section-4}) the lifetime of the 
DM is much longer than the age of the Universe. As a result it could explain the observed positron anomalies at PAMELA~\cite{PAMELA_positron,PAMELA_ep}, 
Fermi~\cite{Fermi_ep,Fermi_positron} and AMS-02~\cite{ams,ams_ep} without conflicting with the antiproton data.

\section{Co-genesis of Visible and Dark Matter} \label{section-3}

\subsection{Baryon asymmetry}

\begin{figure}
\begin{center}
\epsfig{file=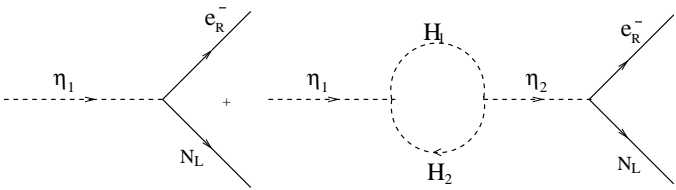, width=0.5\textwidth}
\caption{The interference of tree-level and self-energy correction diagrams which give rise to CP violation.}
\label{fig-cpviolation}
\end{center}
\end{figure}

In this section we explain the details of simultaneously creating the observed baryon asymmetry and the relic abundance of 
DM in our model. We assume that the hidden sector inflaton $\phi$ with mass $m_{\phi}$ decays into the SM 
degrees of freedom through $\eta$ and $\chi$ as depicted in Fig.~\ref{fig-1}. We further assume this gives 
rise to a reheat temperature:
\begin{equation}
T_R \sim 0.1 \sqrt{\Gamma_\phi M_{\rm Pl}} \gae 100 {\rm GeV}\,.
\end{equation}
To generate baryon asymmetry we need CP violation for which we assume that there exist two $\eta$ fields: $\eta_1$ and $\eta_2$ 
of masses $M_1$ and $M_2$. Since their couplings with $N_L$ and $\ell_R$ are in general complex, the $B-L$ conserving decay of 
the lightest one can give rise to CP violation through the interference of tree level and self energy correction diagrams as shown 
in the Fig.~\ref{fig-cpviolation}. 
The CP violation due to the decay of the lightest $\eta$ can be estimated to be~\cite{ma&sarkar},
\begin{equation}
\epsilon_L=\frac{ {\mathrm Im}\left[ (\mu_1 \mu_2^*) \sum_{\alpha \beta} h^1_{\alpha \beta}
h^{2*}_{\alpha \beta} \right]}{16 \pi^2 (M_2^2-M_1^2)} \left[ \frac{M_1}
{\Gamma_1}\right]=-\epsilon_{N_L}\,,
\label{cpasymmetry}
\end{equation}
where 
\begin{equation}
\Gamma_1=\frac{1}{8 \pi  M_1}\left( \mu_1 \mu_1^* + M_1^2 
\sum_{i,j} h^1_{\alpha \beta} h^{1*}_{\alpha \beta} \right)\,.
\label{eta_decayrate}
\end{equation} 
Now assuming $\mu_1\sim \mu_2\sim M_1 \sim M_2 $ and $h^1_{\alpha \beta} \sim h^2_{\alpha \beta}\sim {\cal O}( 10^{-2})$ 
we get from Eqs. (\ref{cpasymmetry}) and (\ref{eta_decayrate}) the CP asymmetry $|\epsilon_L|=|\epsilon_{N_L}| 
\simeq 10^{-5}$.

Since the decay of the lightest $\eta$ does not violate lepton number, so it can not produce a 
net $B-L$ asymmetry. But it will produce an {\it equal and opposite} $B-L$ asymmetry between $N_L$ and 
$\ell_R$~\cite{dirac_lep,sahu&sarkar,sahu&mcdonald}. The two asymmetries, which remain isolated from 
each other before electroweak phase transition, can be given by: 
\begin{equation}\label{non-thermal-asymmetry}
{\mathcal Y}_{\rm B-L} = B_\eta \epsilon_L \frac{n_\phi}{ s}|_{T=T_R}=-{\mathcal Y}^{\rm asy}_{N_L}
\end{equation}
where $n_\phi=\rho_\phi/m_\phi$ is the inflaton density and $s=(2\pi^2/45)g_* T^3$ is the entropy density. 
The branching fraction in the above equation is defined by:
\begin{equation}
B_\eta= \frac{ \Gamma(\phi\to \eta^+\eta^-) } {\Gamma (\phi \to {\rm all})}\,.
\end{equation} 
Using $\rho_\phi|_{T=T_R}= (\pi^2/30) g_* T_R^4$ in Eq.~(\ref{non-thermal-asymmetry}) we get
\begin{equation} 
{\mathcal Y}_{\rm B-L}=\frac{3}{4}B_\eta \epsilon_L \frac{T_R}{m_\phi}= -{\mathcal Y}^{\rm asy}_{N_L}\,.
\end{equation}
The $B-L$ asymmetry in $\ell_R$ can be transformed to 
$\ell_L$ through the lepton number conserving process: $\ell_R \ell^c_R \leftrightarrow \ell_L \ell^c_L$ mediated 
via the SM Higgs as it remains equilibrium above electroweak phase transition. As a result the $B-L$ asymmetry in 
the lepton sector can be converted to baryon asymmetry through the $SU(2)_L$ sphalerons while leaving an equal and 
opposite $B-L$ asymmetry in $N_L$. The conversion of $B-L$ asymmetry to the baryon asymmetry is obtained by :
\begin{equation}
{\mathcal Y}_B=\frac{24}{92}B_\eta \epsilon_L \frac{T_R}{m_\phi}\,.
\end{equation}
For $T_R/m_\phi\approx 10^{-4}$ and $\epsilon_L\approx 10^{-5}$, we can achieve the observed baryon asymmetry $Y_B\approx 
{\cal O}(10^{-10})$. This leads to the DM to baryon abundance:
\begin{equation}
\frac{ {\mathcal Y}^{\rm asy}_{N_L} }{{\mathcal Y}_B} = \frac{92}{32}\,.
\label{dmtobaryon}
\end{equation}
A crucial point to note here is that the asymmetric component of DM and baryon asymmetry are 
produced by a non-thermal decay of the $\phi$ decay products, $\eta$ and $\chi$. An obvious danger 
of washing out this asymmetry comes from the $B-L$ violating process $N_L \ell_R \rightarrow \ell_L \ell_{L}$ 
through the mixing between $\eta$ and $\chi$. However, this process is suppressed by a factor 
$(m^2/M_\eta^2 M_\chi^2)^2$ for $m\ll M_\eta, M_\chi$ and hence it cannot compete with the Hubble 
expansion parameter at $T_R \sim 100 $ GeV. Another lepton number violating process is $\ell_L \ell_L \to HH$ 
mdeiated by $S$ and $\psi$. However, the rate of this process: $\Gamma \sim M_\nu^2 T_R^3/\langle H \rangle^4$ 
is much less than the Hubble expansion parameter at $T_R \sim 100 $ GeV. As a result the net $B-L$ asymmetry 
produced by the decay of $\eta$ will be converted to the required baryon asymmetry without suffering any washout.

\subsection{Dark Matter abundance}

Let us now calculate the required DM to baryon ratio: 
\begin{equation}\label{ratio}
\frac{\Omega_{N_L} } {\Omega_{\rm B} }=\frac{ {\mathcal Y}^{\rm asy}_{N_L} }{{\mathcal Y}_B}  \frac{M_{N}}{m_n}\,,
\end{equation}
where $m_n$ is the mass of a nucleon, and $M_N$ is the Majorana mass of the DM candidate $N_L$.

As we discuss in section (\ref{section-4}), $N_L$ mass is required to be ${\cal O}(\rm TeV)$ to explain 
the observed cosmic ray anomalies at PAMELA~\cite{PAMELA_positron,PAMELA_ep}, Fermi~\cite{Fermi_ep,Fermi_positron} and recently 
at AMS-02~\cite{ams,ams_ep}. 
However, for ${\cal O}(\rm TeV)$ mass of $N_L$, Eq.~(\ref{ratio}) gives $\Omega_{N_L} >> \Omega_B$. Fortunately 
this is not be the case, because of the Majorana mass of $N_L$ which give rise to rapid oscillation between 
$N_L$ and $N_L^c$~\cite{oscillation}. As a result the $N_L$ asymmetry can be further reduced through the annihilation 
process: $N_L N_L^c \to Z_{\rm B-L} \to f\bar{f}$, where $f$ is the SM fermion. 

Note that the decay of $\eta$ also give rise to a dominant $B-L$ symmetric abundance of $N_L$ and is given by:
\begin{equation}
{\mathcal Y}^{\rm sym}_{N_L}=\frac{3}{4}B_\eta \frac{T_R}{m_\phi}
\end{equation} 
which is larger than the asymmetric component ${\mathcal Y}^{\rm asy}_{N_L}$ by five orders of magnitude and hence required further 
depletion to match with the observed DM abundance. 

The total $N_L$ abundnace ${\mathcal Y}_{N_L}={\mathcal Y}^{\rm sym}_{N_L}+ {\mathcal Y}^{\rm asy}_{N_L}\approx {\mathcal Y}^{\rm sym}_{N_L}$, 
thus produced non-thermally, can be matched with the observed DM abundance by requiring that the annihilation cross-section:
\begin{equation}
\langle \sigma|v|\rangle_{\rm ann}\equiv
\langle \sigma |v| \rangle_{(N_L \overline{N_L} \to Z_{\rm B-L}\to \sum_f f\bar{f})} \approx \frac{1}{4\pi} \frac{M_N^2}{v_{\rm B-L}^4}\,,
\label{cross-section}
\end{equation}
is larger than the freeze-out value $\langle \sigma|v|\rangle_F =2.6 \times 10^{-9} {\rm GeV}^{-2}$. Note that in the above equation 
we have used the mass of $Z_{\rm B-L}$ boson to be:
\begin{equation}
M_{Z'}= g_{\rm B-L} v_{\rm B-L}\,,
\end{equation}
with $v_{\rm B-L}$ is the $B-L$ symmetry breaking scale. In an expanding Universe, the annihilation cross-section (\ref{cross-section}) 
has to compete with the Hubble expansion parameter: 
\begin{equation}
H=1.67 g_*^{1/2} \frac{T^2}{M_{\rm pl}}\,,
\end{equation}
and the details of dynamics can be obtained by solving the relevant Boltzmann equations:
\begin{eqnarray}
\frac{dn_\eta}{dt}+3 n_\eta H=-\Gamma_\eta n_\eta \,,\nonumber\\
\frac{dn_{N_{L}}}{dt} + 3 n_{N_{L}} H = -\langle \sigma|v| \rangle_{\rm ann}
n_{N_{L}}^2 + \Gamma_\eta n_\eta\,.
\label{eq:boltzmann}
\end{eqnarray}
If we omit the production term from the thermal bath, i.e., $\Gamma_\eta n_\eta \to 0$ in Eq. (\ref{eq:boltzmann}), then 
$\frac{dn_{N_{L}}}{dt} << 3n_{N_{L}} H$. In this approximation we obtain,
\begin{equation}
{\mathcal Y}_{N_L} \equiv \frac{n_{N_L}}{s} \simeq  \frac{3 H}{\langle \sigma|v|\rangle_{\rm ann}  s} \,,
\label{eq:Yfreezeout}
\end{equation}
where $s$ is the entropy density. In the above equation ${\mathcal Y}_{N_L}$ has to be matched with the observed DM abundance: 
\begin{equation}
\left( {\mathcal Y}_{N_L} \right)_{\rm obs} 
= 4 \times 10^{-13} \left(\frac{1 \TeV}{M_{N}} \right) \left(
\frac{\Omega_{\rm DM} h^2}{0.11}\right)\,.
\label{eq:obsDM}
\end{equation}
The matching of Eqs. (\ref{eq:Yfreezeout}) and (\ref{eq:obsDM}) at $T=T_R$, gives a constraint on the annihilation 
cross-section to be: 
\begin{eqnarray}
\frac{ \langle \sigma |v| \rangle_{\rm ann} }{ \langle \sigma|v|\rangle_F} &=& 2.74
\left(\frac{M_N}{3 \TeV} \right) \left(\frac{0.11}{\Omega_{\rm DM}h^2} \right)\nonumber\\
&&  \left( \frac{100 {\rm GeV}}{T_R} \right)\,.
\label{enhanced-annihilation}
\end{eqnarray}
The above equation implies that the annihilation cross-section (\ref{cross-section}) is a few times larger than the 
freeze-out value for a reheat temperature of 100 GeV. Now combining Eqs. (\ref{cross-section}) and (\ref{enhanced-annihilation}) 
we can get a constraint on the $B-L$ breaking scale to be
\begin{eqnarray}
v_{\rm B-L} = 3.16 \TeV
&&\left( \frac{\Omega_{\rm DM} h^2 }{0.11} \right)^{1/4}
 \left( \frac{M_N}{3 \TeV} \right)^{1/4}\, \nonumber \\
&&\times \left(\frac{T_R}{100 \GeV} \right)^{1/4}\,.
\end{eqnarray}

\noindent\section{Decaying DM and Cosmic Ray Anomalies} \label{section-4} 

The lepton number is violated through the mixing between $\eta$ and $\chi$ as defined 
by $m^2 \eta^\dagger \chi$. Therefore, the lightest $N_L$, which is the candidate of DM, is not stable. We 
assume that $m << M_\eta, M_\chi$. This gives a suppression in the decay rate of DM. In other words the lifetime 
of DM is longer than the age of the Universe. The only available channel for the decay of lightest $N_L$ is three body decay: 
\begin{equation}
N_L \rightarrow e_{\alpha R}^- e_{\beta L}^+ \overline{\nu}_{\gamma L}\,,
\end{equation} 
with $\beta \neq \gamma$. Since the coupling of $\chi$ to two lepton doublets is antisymmetric, i.e., 
$\beta \neq \gamma$, the decay of $N_L$ is not necessarily to be flavor conserving. In particular the decay 
mode: $N_L \rightarrow \tau_{R}^- \tau_{L}^+ \overline{\nu}_{e L} (\overline{\nu}_{\mu L})$, violates $L_e$ ($L_\mu$) 
by one unit while it violates $L=L_e + L_\mu + L_\tau$ by two units.

\begin{figure}[htbp]
\begin{center}
\epsfig{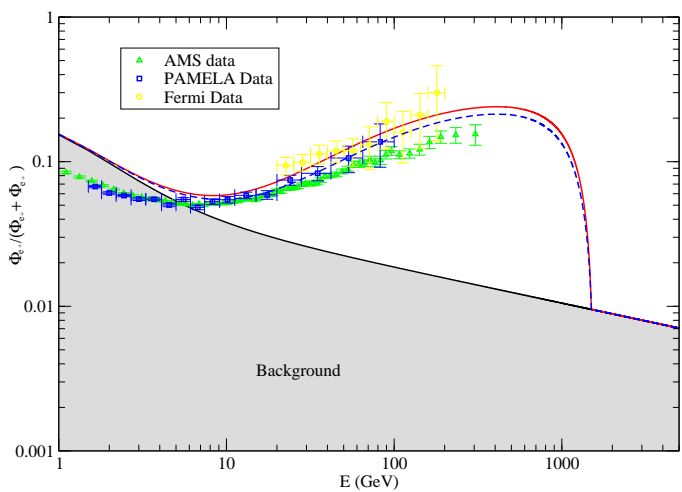}
\caption{Positron excess from lightest $N_L \rightarrow \tau^- \tau^+ \bar{\nu} $ with $M_N=3$ TeV. The 
red-solid (top) and Blue-dashed (bottom) lines are shown for $\tau_N=4\times 
10^{25} {\rm sec}$ and $\tau_N=5\times 10^{25} {\rm sec}$ respectively. The fragmentation function has been 
calculated using PYTHIA~\cite{Sjostrand:2006za}.}
\label{fig-positron}
\end{center}
\end{figure}

\begin{figure}[htbp]
\begin{center}
\epsfig{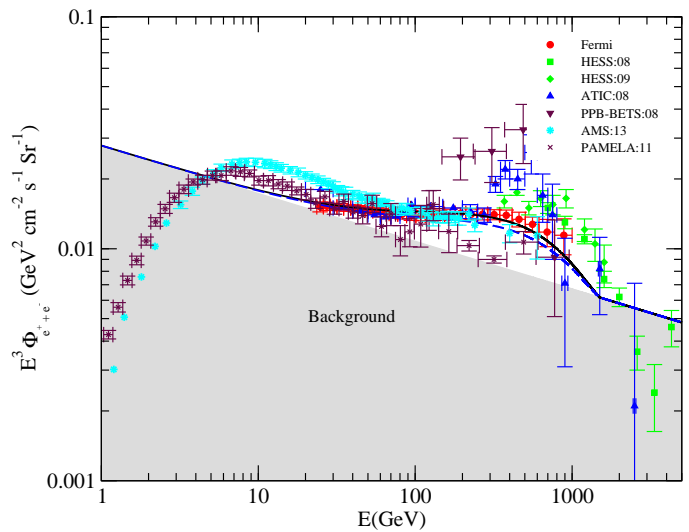}
\caption{Total electron plus positron flux from lightest $N_L \rightarrow \tau^- \tau^+ \bar{\nu} $ with $M_N=3$ TeV. The 
Black-solid (top) and Blue-dashed (bottom) lines are shown for $\tau_N=4\times 10^{25} {\rm sec}$ and $\tau_N=5\times 10^{25} 
{\rm sec}$ respectively. The fragmentation function has been calculated using PYTHIA~\cite{Sjostrand:2006za}.}
\label{fig-ep}
\end{center}
\end{figure}

In the mass basis of $N_L$ the lifetime can be estimated to be 
\begin{eqnarray}
\tau_{N}  &=& 8.0 \times 10^{25} {\rm s} \left( \frac{10^{-2}}{h}\right)^2 \left( \frac{10^{-8.5}}{f} 
\right)^2 \nonumber\\
&& \left(\frac{50 \GeV}{m} \right)^4 \left( \frac{m_\phi}{10^6 \GeV} \right)^8 
\left( \frac{3 \TeV}{M_N}\right)^5\,,
\end{eqnarray}
where we assume that $M_\eta \simeq M_\chi \approx m_\phi$ in order to get a lower limit on the lifetime of $N_L$. 
The prolonged lifetime of $N_L$ may explain the current cosmic ray anomalies observed by PAMELA~\cite{PAMELA_positron,PAMELA_ep}, 
Fermi~\cite{Fermi_ep,Fermi_positron} and recently at AMS-02~\cite{ams,ams_ep}. The electron and positron energy spectrum can be estimated by using the
same set-up as in Ref.~\cite{decay}.  In Figs.~\ref{fig-positron} and ~\ref{fig-ep} we have shown the integrated electron and 
positron fluxes in a typical decay mode: $N_L \rightarrow \tau^- \tau^+ \bar{\nu} $ up to the maximum available energy $M_N/2$ for 
two values of decay life-time, namely $\tau_N=4\times 10^{25}$ sec and $\tau_N=5\times 10^{25}$ sec. \footnote{The constraints on 
                            the $\tau^+ + \tau^-$ emission modes by gamma-ray emissions from the Galactic center and dwarf spheroidals 
                            within the Galaxy depends on the density profile. Since we adopt a cored profile, the constraints are much 
                            weaker than those from the Galactic center and dwarf spheroidals~\cite{Yuan:2013eja}}.
From there it can be seen that the decay of $N_L$ can nicely 
explain the observed cosmic ray excesses at PAMELA, Fermi and at AMS-02.  While doing so we assume that the branching 
fraction in the decay of $N_L$ to $\tau^-\tau^+ \bar{\nu}$ is significantly larger than the other viable decay
modes: $N_L \rightarrow \mu^- \mu^+ \bar{\nu} $ and $N_L \rightarrow e^- e^+ \bar{\nu} $. 

Another potential signature of this scenario is the emission of energetic neutrinos from 
the Galactic center~\cite{GCnu} which can be checked by future experiments such as IceCube 
DeepCore~\cite{Cowen:2008zz} and KM3NeT~\cite{Kappes:2007ci}.

\section{Direct detection of dark matter and constraints} \label{section-5}

The interaction of $N_L$ on the nucleons can give rise to a coherent spin-independent elastic 
scattering, mediated by the $Z_{\rm B-L}$ gauge boson,  through $t$-channel process. In the limit 
of zero-momentum transfer the resulting cross-section is given by: 
\begin{eqnarray}
\sigma_{N_L n}  &= & \frac{\mu^2_{N_L n}}{64 \pi v_{\rm B-L}^4}\left(Y_{\rm B-L}^q Y_{\rm B-L}^{N_L}\right)^2 \nonumber\\
&& \left(Z \frac{f_p}{f_n}  + (A-Z) \right)^2 f_n^2  
\label{dm-n}
\end{eqnarray}  
where $f_n$ and $f_p$ introduces the hadronic uncertainties in the elastic cross-section and $\mu_{N_L n}$ is the 
reduced mass of DM-nucleon system, given by
\begin{equation}
\mu_{N_L n}=\frac{M_N m_n }{M_N + m_n}\,.
\end{equation}
Since $M_N >> m_n$, one gets $ \mu_{N_L n} \approx m_n$. In Eq. (\ref{dm-n}), the symbols $Y_{\rm B-L}^q$ and 
$Y_{\rm B-L}^{N_L}$ represent $B-L$ charge of quark and $N_L$ respectively. The value of $f_n$ vary within a wide 
range: $0.14 < f_n < 0.66$, as quoted in ref.~\cite{fn-value}. Here after we take $f_n\simeq \frac{1}{3}$, the central value. 

\begin{figure}[htbp]
\begin{center}
\epsfig{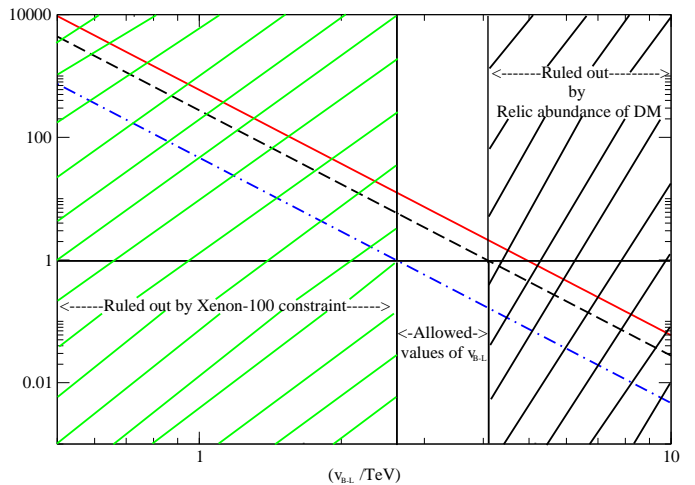}
\caption{$\langle \sigma|v|\rangle_{\rm ann}/\langle \sigma|v|\rangle_F$, shown by dashed black and $\sigma_{\rm DMn}/\sigma_{\rm xenon100}$ shown 
by solid red (iso-spin conserving) and blue dot-dashed (iso-spin violating) as function of $v_{\rm B-L}$ for a typical value of the 
DM mass: $M_N=3 \TeV$.  
} 
\label{comparison}
\end{center}
\end{figure}

At present the strongest constraint on spin-independent DM-nucleon cross-section is given by Xenon-100, which 
assumes $f_p/f_n=1$ with $Z=54$, while $A$ varies between 74 to 80. This is the case of iso-spin conserving case. 
For a 3 TeV DM, Xenon-100 gives an upper bound on the DM-nucleon cross-section to be  $\sigma_{N_L n} < {\cal O}( 10^{-43}) 
{\rm cm}^{2}$ at 90\% confidence level~\cite{Aprile:2012nq}. From Eq.(\ref{dm-n}) we can estimate the DM-nucleon cross-section: 
\begin{eqnarray}
 \sigma_{N_L n}  &=& 2.15 \times 10^{-43} {\rm cm}^{2} \left(\frac{\mu_{N_L n}}{\rm GeV} \right)^2 \nonumber\\
&& \left( \frac{5 \TeV}{v_{\rm B-L}}\right)^4\,.
\end{eqnarray}
Thus the $\sigma_{N_L n}$ cross-section is in the right order of magnitude and it is compatible with the latest Xenon-100 
limit~\cite{Aprile:2012nq}. However, from Eq. (\ref{cross-section}) we see that for $v_{\rm B-L}=5 \TeV$ and $M_N=3 \TeV$, the 
annihilation cross-section: $\langle \sigma|v|\rangle_{\rm ann} < \langle \sigma|v|\rangle_F = 2.6 \times 10^{-9} {\rm GeV}^{-2}$. 
This implies that we get DM abundance more than the observed value and hence $v_{\rm B-L} \geq 5 \TeV$ 
is not allowed. On the other hand, for $v_{\rm B-L} < 5 \TeV$ we can get right amount of DM abundance. But those values of $v_{\rm B-L}$ 
are not allowed by Xenon-100 constraint as they give large DM-nucleon cross-section. These features can be easily read from Fig.~\ref{comparison}, 
where we have shown the compatibilty of $B-L$ breaking scale with relic abundance (dashed black line) and direct detection constraint 
(solid red for iso-spin conserving and dot-dashed blue for iso-spin violating) from Xenon-100.    

From Eqs.(\ref{cross-section}) and (\ref{dm-n}) we see that both the cross-sections: $\langle \sigma |v| \rangle_{\rm ann}$ 
and $\sigma_{N_L n}$ vary inversely as $4^{\rm th}$ power of $B-L$ breaking scale. Therefore, we need large $\langle \sigma |v| \rangle_{\rm ann}$ 
to get the right amount of relic abundance of DM, while small $\sigma_{N_L n}$ is required to be compatible with the direct detection 
limits from Xenon-100. In other words, we need small $v_{\rm B-L}$ to get the right amount of relic abundance, while large $v_{\rm B-L}$ 
is required to be compatible with the direct detection limits. 

From Fig.~\ref{comparison}, we see that for iso-spin conserving case (solid red line) we don't get any value of $v_{\rm B-L}$, which is compatible with 
the relic abundance and the direct detection constraint on DM. However, this constraints can be evaded by considering an iso-spin violating 
DM-nucleon interaction~\cite{isospin_violation} as shown in the Fig.~\ref{comparison} by dot-dashed blue line. From there we see that a small window of $B-L$ 
breaking scale: $v_{\rm B-L}$= (2.5 TeV  - 4 TeV) can give $\langle \sigma|v|\rangle_{\rm ann} \gsim \langle \sigma|v|\rangle_F$ 
and $\sigma_{N_L n} < \sigma_{\rm Xenon100}$ for $M_N=3 \TeV$.    

Thus we saw that the DM satisfy the direct detection constraints from Xenon-100 only in case of iso-spin violation and within a small 
window of $B-L$ breaking scale:  $v_{\rm B-L}$= (2.5 TeV  - 4 TeV). It is worth mentioning that the model though involves many 
parameters to explain the cosmic ray anomalies from decaying DM, but the relic abundance and the compatibility with direct detection 
constraints of the latter involves a single parameter, i.e. the $B-L$ breaking scale: $v_{\rm B-L}$. In one hand, if $v_{\rm B-L} > 4$ TeV, 
then the annihilation cross-section of DM is smaller than the freeze-out value (see Eq. \ref{cross-section}) and hence the model produces 
large DM abundance. On the other hand, if $v_{\rm B-L} < 2$ TeV, then the DM doesn't satisfy the direct detection constraints from 
Xenon-100 (see Eq. (\ref{dm-n}). Note that the above conclusions are independent of other parameters involved in explaining cosmic ray 
anomalies and baryon asymmetry. Therefore, our scenario is strongly constrained in terms of the model parameter and can be checked at the 
future terrestrial experiments such as Xenon-1T.        

\section{Conclusions} \label{section-6}
 
We studied a non-thermal scenario in a gauged $B-L$ extension of the SM to explain a common origin behind 
DM abundance and baryon asymmetry. The $B-L$ symmetry is broken at a TeV scale which gives a Majorana mass to the DM,
while the baryon asymmetry is created via lepton number conserving 
leptogenesis mechanism and therefore it does not depend on the $B-L$ breaking scale. 
Since the lepton number is violated, the DM is no longer stable and slowly decays into the lepton sector as it 
carries a net leptonic charge. Since the decay rate of DM is extremely slow, it could explain the positron excess 
observed at PAMELA, Fermi and recently at AMS-02 without conflicting with the antiproton data.

We also checked the 
compatibility of a TeV scale DM with the spin-independent DM-nucleon scattering at Xenon-100, which at present gives the strongest constraint 
on DM-nucleon cross-section. We have found that in the case of iso-spin conserving, the spin independent DM-nucleon cross-section is incompatible with 
the relic abundance of DM. On the other hand, by assuming the iso-spin violation interaction, we found a small window of $B-L$ breaking 
scale: $v_{\rm B-L}$= (2.5 TeV  - 4 TeV), which can yield right amount of DM abundance while explaining the positron excess.
 This implies the corresponding $B-L$ gauge boson ({\it i.e.} $Z'$-gauge boson) is necessarily to be 
at a TeV scale which can be searched at the LHC.

\section{Acknowledgements} We thank Anupam Mazumdar, Kazunori Nakayama, Chiara Arina and Julian Heeck for useful discussions. KK is 
supported in part by Grant-in-Aid for Scientific research from the Ministry of Education, Science, Sports, and Culture (MEXT), 
Japan, No. 21111006, No. 22244030, and No. 23540327. NS is partially supported by the Department of Science and Technology Grant 
SR/FTP/PS-209/2011. 
 

\end{document}